\documentclass[12pt]{article}

\usepackage{amsmath,amssymb}
\usepackage{bm}
\usepackage{mathrsfs}
\usepackage{multirow}
\usepackage{float}
\usepackage[usenames]{color}
\usepackage{indentfirst}  
\usepackage{cite}

\allowdisplaybreaks[4]

\newcommand{{\Slashp}}{p\!\!\!\!\!\big/}
\newcommand{{\Slashq}}{q\!\!\!\!\!\big/}

\begin{document}

\title{Predictions of $SU(9)$ orbifold family unification}

\author{
Yuhei \textsc{Goto}\footnote{E-mail: 14st302a@shinshu-u.ac.jp} ~~and~ 
Yoshiharu \textsc{Kawamura}\footnote{E-mail: haru@azusa.shinshu-u.ac.jp}\\
{\it Department of Physics, Shinshu University, }\\
{\it Matsumoto 390-8621, Japan}
}


\maketitle
\begin{abstract}
We study predictions of orbifold family unification models
with $SU(9)$ gauge group on a six-dimensional space-time
including the orbifold $T^2/Z_2$,
and obtain relations among sfermion masses 
in the supersymmetric extension of models.
The models have an excellent feature
that just three families of the standard model fermions
exist in a pair of Weyl fermions in the {\bf 84} representation
as four-dimensional zero modes, 
without accompanying any mirror particles.
\end{abstract}

\section{Introduction}
\label{sec:Intro}

Gauge theories on a higher-dimensional space-time
including an orbifold as an extra space possess suitable properties 
to realize a family unification~\cite{BB&K,W&T,GM&N1,GLM&N,GLM&S,
GM&N2,KK&O,M&N,K&M1,GK&M,FKMN&S,GK&M2,AFKMN&S,FKMNS&T}.
Three families of the standard model (SM) fermions
are embedded into a few multiplets of a large gauge group.
Extra fermions including mirror particles can be eliminated
and the SM fermions can survive as zero modes,
through the orbifold breaking mechanism. 

In our previous work, we have found a lot of possibilities
that three families of the SM fermions appear
as zero modes from a pair of Weyl fermions
in $SU(N)$ gauge theories ($N=9 \sim 13$)
on the six-dimensional (6D) space-time
$M^4 \times T^2/Z_M$ ($M=2, 3, 4, 6$)~\cite{GK&M}.
The subjects left behind are to construct realistic models
and to find out model-dependent predictions.

In this paper, we focus on the orbifold family unification
in the minimal setup,
because models tend to be more complex and less realistic
by extending the structure of space-time
and/or the ingredients of models such as gauge symmetries 
and representations of matters.
We take orbifold family unification models
based on $SU(9)$ gauge symmetry 
on $M^4 \times T^2/Z_2$ as a starting point,
examine the reality of models, 
and find out some predictions.
For the reality, we use the appearance of Yukawa interactions
from interactions in the 6D bulk as a selection rule.
For the predictions, we search specific relations
among sfermion masses 
in the supersymmetric (SUSY) extension of models.

The contents of this paper are as follows. 
In Sec.$\ref{sec:SU9}$, we explain our setup and its properties.
In Sec.$\ref{Sec:Predictions}$, we carry out 
the examination for the reality of models 
and the search for predictions. 
Sec.$\ref{C&D}$ is devoted to conclusions and discussions.

\section{$SU(9)$ orbifold family unification}
\label{sec:SU9}

\subsection{Setup}

Our space-time is assumed to be the product of
four-dimensional (4D) Minkowski space-time $M^4$
and two-dimensional (2D) orbifold $T^2/Z_2$.
The $T^2/Z_2$ is obtained by dividing 2D lattice $T^2$
by the $Z_2$ transformation: $z \rightarrow -z$,
where $z$ is a complex coordinate.
Then, $z$ is identified with $(-1)^l z + m e_1 + n e_2$,
where $l$, $m$ and $n$ are integers,
and $e_1$ and $e_2$ are basis vectors of $T^2$.

We impose the following boundary conditions (BCs) 
on a 6D field $\Psi(x,z)$,
\begin{align}
  \Psi(x,-z) &= T_{\Psi}[P_0]\Psi(x,z), \\
  \Psi(x, e_1-z) &= T_{\Psi}[P_1]\Psi(x,z), \\
  \Psi(x, e_2-z) &= T_{\Psi}[P_2]\Psi(x,z),
\end{align}
where $T_{\Psi}[P_0]$, $T_{\Psi}[P_1]$ and $T_{\Psi}[P_2]$ 
represent the representation matrices, and
$P_0$, $P_1$ and $P_2$ stand for 
the representation matrices of $Z_2$ transformations 
$z \rightarrow -z$, $z \rightarrow e_1-z$ and $z \rightarrow e_2-z$ 
for fields with the fundamental representation. 

The eigenvalues of $T_{\Psi}[P_0]$, $T_{\Psi}[P_1]$ 
and $T_{\Psi}[P_2]$ are interpreted 
as the $Z_2$ parities on $T^2/Z_2$. 
The fields with even $Z_2$ parities have zero modes. 
Here, zero modes mean 4D massless fields 
surviving after compactification. 
Massive modes are called $\lq\lq$Kaluza-Klein modes'',
and they do not appear in the low-energy world
because they have heavy masses 
of $O(1/L)$ where $L$ is the size of extra space.
Fields including an odd $Z_2$ parity do not have zero modes.
Hence, the reduction of symmetry occurs upon compactification,
unless all components of multiplet have common $Z_2$ parities. 
This type of symmetry breaking mechanism is called
the $\lq\lq$orbifold breaking mechanism''.\footnote{
The $Z_2$ orbifolding was used in superstring theory~\cite{A} 
and heterotic $M$-theory~\cite{H&W1,H&W2}.
In field theoretical models, it was applied to the reduction of global SUSY~\cite{M&P,P&Q}, which is
an orbifold version of Scherk-Schwarz mechanism~\cite{S&S,S&S2}, and then
to the reduction of gauge symmetry~\cite{K1}.
}

We start with 6D $SU(9)$ gauge theories
containing a pair of Weyl fermions $(\Psi_+, \Psi_-)$.
These fermions own a same representation of $SU(9)$
but different chiralities and are represented as
\begin{align}
  \Psi_+ &= \frac{1+\Gamma_7}{2}\Psi = \left(\begin{array}{cc} \frac{1-\gamma_5}{2}&0\\0&\frac{1+\gamma_5}{2} \end{array} \right)
\left( \begin{array}{c} \psi^1 \\ \psi^2 \end{array}\right) 
= \left( \begin{array}{c} \psi^1_L \\ \psi^2_R \end{array}\right),  \label{eq:6weyl+} \\
  \Psi_- &= \frac{1-\Gamma_7}{2}\Psi = \left(\begin{array}{cc} \frac{1+\gamma_5}{2}&0\\0&\frac{1-\gamma_5}{2} \end{array} \right)
\left( \begin{array}{c} \psi^1 \\ \psi^2 \end{array}\right) 
= \left( \begin{array}{c} \psi^1_R \\ \psi^2_L \end{array}\right),  \label{eq:6weyl-}
\end{align} 
where $\Psi_+$ and $\Psi_-$ are fermions 
with positive and negative chirality, respectively, 
and $\Gamma_7$ and $\gamma_5$ are the chirality operators 
for 6D fermions and 4D fermions, respectively.
We use the following representation for the $8 \times 8$
gamma matrices $\Gamma^M$ $(M=0,1,2,3,5,6)$, 
\begin{align}
\Gamma^{\mu}=\gamma^{\mu}\otimes\sigma^3,~~
\Gamma^5=I_{4\times4}\otimes i\sigma^1,~~
\Gamma^6=I_{4\times4}\otimes i\sigma^2,
\label{Gammas}
\end{align}
where $\gamma^{\mu}$ $(\mu =0,1,2,3)$, $I_{4\times4}$
and $\sigma^i$ $(i=1,2,3)$
are the $4 \times 4$ gamma matrices, the $4 \times 4$
unit matrix and Pauli matrices, respectively.
In the chiral representation, $\gamma^{\mu}$ are given by
\begin{align}
\gamma^{\mu}=
\left(\begin{array}{cc} 0&\sigma^{\mu}\\
\overline{\sigma}^{\mu} & 0
\end{array} \right),
\label{gammas}
\end{align}
where $\sigma^{\mu} = (I, \bm{\sigma})$ and 
$\overline{\sigma}^{\mu} = (I, -\bm{\sigma})$.
Here, $I$ is the $2 \times 2$ unit matrix.
The $\Gamma^M$ satisfy the Clifford algebra 
$\{\Gamma^M,\Gamma^N\}=2\eta^{MN}$.
Note that the theories are free of chiral anomalies,
as a result of cancellations between contributions from $\Psi_+$
and those from $\Psi_-$.

When we take the representation matrices
\begin{align}
  P_0 &= \text{diag}([+1]_{p_1},[+1]_{p_2},[+1]_{p_3},[+1]_{p_4},[-1]_{p_5},[-1]_{p_6},[-1]_{p_7},[-1]_{p_8}), \notag\\
  P_1 &= \text{diag}([+1]_{p_1},[+1]_{p_2},[-1]_{p_3},[-1]_{p_4},[+1]_{p_5},[+1]_{p_6},[-1]_{p_7},[-1]_{p_8}), \label{eq:RM}\\
  P_2 &= \text{diag}([+1]_{p_1},[-1]_{p_2},[+1]_{p_3},[-1]_{p_4},[+1]_{p_5},[-1]_{p_6},[+1]_{p_7},[-1]_{p_8}), \notag
\end{align}
the breakdown of $SU(9)$ gauge symmetry occurs as
\begin{align}
  SU(9) \rightarrow SU(p_1) \times SU(p_2) \times \cdots \times SU(p_8) \times U(1)^{7-m}, \label{eq:SB}
\end{align}
where $[\pm1]_{p_i}$ represents $\pm1$ for all $p_i$ elements, $p_1+p_2+\cdots+p_8=9$, and
$m$ is a sum of the number of $SU(1)$ and $SU(0)$. 
Here, $SU(1)$ unconventionally stand for $U(1)$ 
and $SU(0)$ means nothing.
Then, 6D fields with the rank-$k$ completely asymmetric 
tensor representation ${9 \choose k}$ are decomposed as
\begin{align}
  {9 \choose k} = \sum_{l_1=0}^{k}\sum_{l_2=0}^{k-l_1} \cdots 
\sum_{l_7=0}^{k-l_1-\cdots-l_6} \left({p_1 \choose l_1}, {p_2 \choose l_2}, 
\cdots , {p_7 \choose l_7}, {p_8 \choose l_8} \right), \label{eq:DECOM}
\end{align} 
where ${a \choose b}$ stands for the representation
whose dimension is the combinatorial number,
$l_8=k-l_1-l_2-\cdots-l_7$ and $p_8=9-p_1-p_2-\cdots-p_7$.

In case that $\Psi_+$ and $\Psi_-$ have ${9 \choose k}$ of $SU(9)$,
we denote the intrinsic $Z_2$ parities of
$\psi_L^1$ and $\psi_L^2$ as 
$(\eta^0_{k+}, \eta^1_{k+}, \eta^2_{k+})$
and $(\eta^0_{k-}, \eta^1_{k-}, \eta^2_{k-})$, respectively.
Then, those of $\psi_R^2$ and $\psi_R^1$ are fixed as 
$(-\eta^0_{k+}, -\eta^1_{k+}, -\eta^2_{k+})$
and $(-\eta^0_{k-}, -\eta^1_{k-}, -\eta^2_{k-})$
from the $Z_2$ invariance of kinetic terms 
and the transformation properties of the covariant derivatives 
$Z_2:D_z \rightarrow -D_z$ and $D_{\bar{z}} \rightarrow-D_{\bar{z}}$.
On the breakdown of $SU(9)$ due to (\ref{eq:RM}),
the $Z_2$ parities of the component with
the representation 
$\left( {p_1 \choose l_1}, {p_2 \choose l_2}, 
\cdots , {p_7 \choose l_7}, {p_8 \choose l_8} \right)$
are given by
\begin{align}
  \mathcal{P}_{0\pm} &= (-1)^{l_5+l_6+l_7+l_8}\eta^0_{k\pm} 
= (-1)^{k-l_1-l_2-l_3-l_4}\eta^0_{k\pm}, \\
  \mathcal{P}_{1\pm} &= (-1)^{l_3+l_4+l_7+l_8}\eta^1_{k\pm} 
= (-1)^{k-l_1-l_2-l_5-l_6}\eta^1_{k\pm}, \\
  \mathcal{P}_{2\pm} &= (-1)^{l_2+l_4+l_6+l_8}\eta^2_{k\pm} 
= (-1)^{k-l_1-l_3-l_5-l_7}\eta^2_{k\pm},
\end{align}
where $\mathcal{P}_{a+}$ and $\mathcal{P}_{a-}$ $(a=0,1,2)$
are the $Z_2$ parities of $\psi_L^1$ and $\psi_L^2$, respectively.
Those of $\psi_R^2$ and $\psi_R^1$ are 
$-\mathcal{P}_{a+}$ and $-\mathcal{P}_{a-}$.

We have found 32 possibilities
that just three families of the SM fermions survive
as zero modes from a pair of Weyl fermions 
with the {\bf 84}$(={9 \choose 3})$ representation of $SU(9)$.
For the list of $(p_1, p_2, p_3, p_4, p_5, p_6, p_7, p_8)$
to derive them, see Table VII in \cite{GK&M}.
They are classified into two cases
based on the pattern of gauge symmetry breaking
such that
$SU(9) \rightarrow SU(3)_C \times SU(2)_L \times SU(3)_F 
\times U(1)^3$ and
$SU(9) \rightarrow SU(3)_C \times SU(2)_L \times SU(2)_F 
\times U(1)^4$.
We study how well the three families of fermions in the SM 
are embedded into
$\Psi_+$ and $\Psi_-$, in the following.

\subsection{$SU(9) \rightarrow SU(3)_C \times SU(2)_L 
\times SU(3)_F \times U(1)^3$}

For the case that $p_1 = 3$, $p_2 = 2$, 
either of $p_3$, $p_4$, $p_5$ or $p_6$ is 3
and either of $p_7$ or $p_8$ is 1,
$SU(9)$ is broken down as 
\begin{align}
  SU(9) \rightarrow SU(3)_C \times SU(2)_L \times SU(3)_F \times U(1)_1 \times U(1)_2 \times U(1)_3, \label{eq:GSM}
\end{align}
where $SU(3)_F$ is the gauge group concerning the family of fermions,
$U(1)_1$ belongs to a subgroup of $SU(5)$ 
and is identified with $U(1)_Y$ in the SM,
and others are originated from $SU(9)$ and $SU(4)$ as
\begin{align}
  SU(9) &\supset SU(5) \times SU(4) \times U(1)_2, \\
  SU(4) &\supset SU(3) \times U(1)_3.
\end{align} 

Let us illustrate the survival of three families in the SM,
using two typical BCs.

~~\\
(BC1)~~$(p_1, p_2, p_3, p_4, p_5, p_6, p_7, p_8)=(3, 2, 3, 0, 0, 0, 0, 1)$\\
In this case, {\bf 84} is decomposed into
particles with the SM gauge quantum numbers
and its opposite ones,
and their $U(1)$ charges and $Z_2$ parities 
are listed in Table \ref{Table:[9,3]}.
\begingroup
\renewcommand{\arraystretch}{1.2}
\begin{table}[htbp]
\begin{center}
  \caption{Decomposition of {\bf 84}
for $(p_1, p_2, p_3, p_4, p_5, p_6, p_7, p_8)=(3, 2, 3, 0, 0, 0, 0, 1)$.
}
  \label{Table:[9,3]}
\begin{tabular}{c|c|c|c|c|c|c}
\hline
$\psi_L^{1(2)}$&$\psi_R^{1(2)}$&$SU(3)_C \times SU(2)_L \times SU(3)_F$&$U(1)_1$&$U(1)_2$&$U(1)_3$&$(\mathcal{P}_0,\mathcal{P}_1,\mathcal{P}_2)$ \\
\hline \hline
$(e_R^{\prime})^c$&$e_R$&$\left( {3 \choose 3}, {2 \choose 0}, {3 \choose 0} \right)=({\bf{1}},{\bf{1}},{\bf{1}})$&$-6$&$12$&$0$&$(+\eta^0,+\eta^1,+\eta^2)$ \\
\hline
$q_L^{\prime}$&$(q_L)^c$&$\left( {3 \choose 2}, {2 \choose 1}, {3 \choose 0} \right)=(\overline{\bf{3}},{\bf{2}},{\bf{1}})$&$-1$&$12$&$0$&$(+\eta^0,+\eta^1,-\eta^2)$ \\
\hline
$(u_R^{\prime})^c$&$u_R$&$\left( {3 \choose 1}, {2 \choose 2}, {3 \choose 0} \right)=({\bf{3}},{\bf{1}},{\bf{1}})$&$4$&$12$&$0$&$(+\eta^0,+\eta^1,+\eta^2)$ \\
\hline
$(u_R)^c$&$u_R^{\prime}$&$\left( {3 \choose 2}, {2 \choose 0}, {3 \choose 1} \right)=(\overline{\bf{3}},{\bf{1}},{\bf{3}})$&$-4$&$3$&$1$&$(+\eta^0,-\eta^1,+\eta^2)$ \\
\hline
$(u_R)^c$&$u_R^{\prime}$&$\left( {3 \choose 2}, {2 \choose 0}, {3 \choose 0} \right)=(\overline{\bf{3}},{\bf{1}},{\bf{1}})$&$-4$&$3$&$-3$&$(-\eta^0,-\eta^1,-\eta^2)$ \\
\hline
$q_L$&$(q_L^{\prime})^c$&$\left( {3 \choose 1}, {2 \choose 1}, {3 \choose 1} \right)=({\bf{3}},{\bf{2}},{\bf{3}})$&$1$&$3$&$1$&$(+\eta^0,-\eta^1,-\eta^2)$ \\
\hline
$q_L$&$(q_L^{\prime})^c$&$\left( {3 \choose 1}, {2 \choose 1}, {3 \choose 0} \right)=({\bf{3}},{\bf{2}},{\bf{1}})$&$1$&$3$&$-3$&$(-\eta^0,-\eta^1,+\eta^2)$ \\
\hline
$(e_R)^c$&$e_R^{\prime}$&$\left( {3 \choose 0}, {2 \choose 2}, {3 \choose 1} \right)=({\bf{1}},{\bf{1}},{\bf{3}})$&$6$&$3$&$1$&$(+\eta^0,-\eta^1,+\eta^2)$ \\
\hline
$(e_R)^c$&$e_R^{\prime}$&$\left( {3 \choose 0}, {2 \choose 2}, {3 \choose 0} \right)=({\bf{1}},{\bf{1}},{\bf{1}})$&$6$&$3$&$-3$&$(-\eta^0,-\eta^1,-\eta^2)$ \\
\hline
$(d_R^{\prime})^c$&$d_R$&$\left( {3 \choose 1}, {2 \choose 0}, {3 \choose 2} \right)=({\bf{3}},{\bf{1}},\overline{\bf{3}})$&$-2$&$-6$&$2$&$(+\eta_0,+\eta_1,+\eta_2)$ \\
\hline
$(d_R^{\prime})^c$&$d_R$&$\left( {3 \choose 1}, {2 \choose 0}, {3 \choose 1} \right)=({\bf{3}},{\bf{1}},{\bf{3}})$&$-2$&$-6$&$-2$&$(-\eta^0,+\eta^1,-\eta^2)$ \\
\hline
$l_L^{\prime}$&$(l_L)^c$&$\left( {3 \choose 0}, {2 \choose 1}, {3 \choose 2} \right)=({\bf{1}},{\bf{2}},\overline{\bf{3}})$&$3$&$-6$&$2$&$(+\eta^0,+\eta^1,-\eta^2)$ \\
\hline
$l_L^{\prime}$&$(l_L)^c$&$\left( {3 \choose 0}, {2 \choose 1}, {3 \choose 1} \right)=({\bf{1}},{\bf{2}},{\bf{3}})$&$3$&$-6$&$-2$&$(-\eta^0,+\eta^1,+\eta^2)$ \\
\hline
$(\nu_R)^c$&$\hat{\nu}_R$&$\left( {3 \choose 0}, {2 \choose 0}, {3 \choose 3} \right)=({\bf{1}},{\bf{1}},{\bf{1}})$&$0$&$-15$&$3$&$(+\eta^0,-\eta^1,+\eta^2)$ \\
\hline
$(\nu_R)^c$&$\hat{\nu}_R$&$\left( {3 \choose 0}, {2 \choose 0}, {3 \choose 2} \right)=({\bf{1}},{\bf{1}},\overline{\bf{3}})$&$0$&$-15$&$-1$&$(-\eta^0,-\eta^1,-\eta^2)$ \\
\hline
\end{tabular}
\end{center}
\end{table}
\endgroup
In the first and second columns, 
particles are denoted by using the symbols in the SM,
and those with primes are regarded as mirror particles.
Here, mirror particles are particles 
with opposite quantum numbers under 
the SM gauge group
$G_{\rm SM} = SU(3)_C \times SU(2)_L \times U(1)_Y$.
The $U(1)$ charges are given up to the normalization.
The $Z_2$ parities of $\psi_L^{1(2)}$ are given
by omitting the subscript $k(=3)$ in the last column.
The $Z_2$ parities of $\psi_R^{2(1)}$ are opposite to
those of $\psi_L^{1(2)}$.

When we assign the intrinsic $Z_2$ parities of
$\psi_L^{1}$ and $\psi_L^{2}$ as
\begin{align}
(\eta_{+}^0,\eta_{+}^1,\eta_{+}^2) = (+1,-1,+1),~~
(\eta_{-}^0,\eta_{-}^1,\eta_{-}^2) = (+1,-1,-1),
\label{BC1-eta}
\end{align}
all mirror particles have an odd $Z_2$ parity
and disappear in the low-energy world.
Then, just three sets of SM fermions 
$(q_L^i, (u_R^i)^c, (d_R^i)^c, l_L^i, (e_R^i)^c)$
survive as zero modes 
and they belong to the following chiral fermions,
\begin{align}
\psi^{1}_L \supset (u_R^i)^c, (e_R^i)^c, (\nu_R)^c,~~
\psi^{2}_R \supset d_R^i,~~
\psi^{1}_R \supset (l_L^i)^c,~
\psi^{2}_L \supset q_L^i,
\label{BC1}
\end{align}
where $i(=1,2,3)$ stands for the family index.
By exchanging $\eta_+^a$ for $\eta_-^a$,
$\psi_L^1$ and $\psi_R^2$ are exchanged
for $\psi_L^2$ and $\psi_R^1$, respectively.
Note that a right-handed neutrino $(\nu_R)^c$ appears alone.
We obtain the same result (\ref{BC1})
by assigning the intrinsic $Z_2$ parities suitably,
in case with $p_4$, $p_5$ or $p_6=3$ in place of $p_3=3$.

~~\\
(BC2)~~$(p_1, p_2, p_3, p_4, p_5, p_6, p_7, p_8)=(3, 2, 3, 0, 0, 0, 1, 0)$\\
In this case, {\bf 84} is decomposed into
particles with the same gauge quantum numbers
but sightly different $Z_2$ parities from those of (BC1).
Concretely, the third $Z_2$ parity $\mathcal{P}_2$ of fields with $l_7 =1$
is opposite to that with $l_8 =1$, i.e., 
$\mathcal{P}_2$ of 
$\left( {3 \choose 2}, {2 \choose 0}, {3 \choose 0} \right)$, 
$\left( {3 \choose 1}, {2 \choose 1}, {3 \choose 0} \right)$, 
$\left( {3 \choose 0}, {2 \choose 2}, {3 \choose 0} \right)$, 
$\left( {3 \choose 1}, {2 \choose 0}, {3 \choose 1} \right)$,
$\left( {3 \choose 0}, {2 \choose 1}, {3 \choose 1} \right)$ and 
$\left( {3 \choose 0}, {2 \choose 0}, {3 \choose 2} \right)$ is given by
$+\eta^2$, $-\eta^2$, $+\eta^2$, $+\eta^2$, $-\eta^2$
and $+\eta^2$, respectively.

Under the same assignment of the intrinsic $Z_2$ parities
as (\ref{BC1-eta}),
all mirror particles have an odd $Z_2$ parity
and disappear in the low-energy world.
Then, just three sets of SM fermions survive as zero modes
such that
\begin{align}
\psi^{1}_L \supset (u_R^i)^c, (e_R^i)^c, (\nu_R)^c,~~
\psi^{2}_R \supset (l_L^i)^c,~~
\psi^{1}_R \supset d_R^i,~~
\psi^{2}_L \supset q_L^i.
\label{BC2}
\end{align}
Note that $(l_L^i)^c$ and $d_R^i$ are embedded into 
$\psi^{2}_R$ and $\psi^{1}_R$, respectively,
different from the case of (BC1).
We obtain the same result (\ref{BC2})
by assigning the intrinsic $Z_2$ parities suitably,
in case with $p_4$, $p_5$ or $p_6=3$ in place of $p_3=3$.\\

We summarize fermions with zero modes
and those gauge quantum numbers in Table \ref{Table:Property1&2}.
Here, $G_{323} = SU(3)_C \times SU(2)_L \times SU(3)_F$,
$l_a$ is a number appearing in a representation ${p_a \choose l_a}$
of $SU(3)_F$ for $a=3, 4, 5$ or $6$,
and, in the 7-th and 8-th columns, the way of embeddings for
the SM species
are shown for $p_8 = 1$ and $p_7 = 1$, respectively.

\begin{table}[htbp]
\begin{center}
  \caption{Gauge quantum numbers of fermions 
with even $Z_2$ parities for $SU(9) \rightarrow G_{323}
\times U(1)_1 \times U(1)_2 \times U(1)_3$.}
  \label{Table:Property1&2}
\begin{tabular}{c|c|c|c|c|c|c|c}
\hline
species& $G_{323}$ &$(l_1,l_2,l_a)$&$U(1)_1$&$U(1)_2$&$U(1)_3$ 
& $p_8 = 1$ & $p_7 =1$ \\
\hline \hline
$q_L^i$&$({\bf{3}},{\bf{2}},{\bf{3}})$&$(1,1,1)$&$1$&$3$&$1$ 
& $\psi_L^{2(1)}$ & $\psi_L^{2(1)}$ \\
\hline
$(u_R^i)^c$&$(\overline{\bf{3}},{\bf{1}},{\bf{3}})$&(2,0,1)&$-4$&$3$&$1$ 
& $\psi_L^{1(2)}$ & $\psi_L^{1(2)}$ \\
\hline
$d_R^i$&$({\bf{3}},{\bf{1}},{\bf{3}})$&$(1,0,1)$&$-2$&$-6$&$-2$ 
& $\psi_R^{2(1)}$ & $\psi_R^{1(2)}$ \\
\hline
$(l_L^i)^c$&$({\bf{1}},{\bf{2}},{\bf{3}})$&$(0,1,1)$&$3$&$-6$&$-2$ 
& $\psi_R^{1(2)}$ & $\psi_R^{2(1)}$ \\
\hline
$(e_R^i)^c$&$({\bf{1}},{\bf{1}},{\bf{3}})$&$(0,2,1)$&$6$&$3$&$1$ 
& $\psi_L^{1(2)}$ & $\psi_L^{1(2)}$ \\
\hline
$(\nu_R)^c$&$({\bf{1}},{\bf{1}},{\bf{1}})$&$(0,0,3)$&$0$&$-15$&$3$
& $\psi_L^{1(2)}$ & $\psi_L^{1(2)}$ \\
\hline
\end{tabular}
\end{center}
\end{table}

\subsection{$SU(9) \rightarrow SU(3)_C \times SU(2)_L 
\times SU(2)_F \times U(1)^4$}

For the case that $p_1 = 3$, $p_2 = 2$, 
either of $(p_3, p_4)$ or $(p_5, p_6)$ is $(2, 1)$ or $(1, 2)$
and either of $p_7$ or $p_8$ is 1,
$SU(9)$ is broken down as 
\begin{align}
  SU(9) \rightarrow SU(3)_C \times SU(2)_L \times SU(2)_F \times U(1)_1 \times U(1)_2 \times U(1)_3 \times U(1)_4 ,
\end{align}
where $U(1)_1$ belongs to a subgroup of $SU(5)$ 
and is identified with $U(1)_Y$ in the SM,
and others are originated from $SU(9)$, $SU(4)$ 
and $SU(3)$ as
\begin{align}
  SU(9) &\supset SU(5) \times SU(4) \times U(1)_2, \\
  SU(4) &\supset SU(3) \times U(1)_3, \\
  SU(3) &\supset SU(2) \times U(1)_4.
\end{align}

The embedding of species are classified into two types,
according to $p_8 =1$ or $p_7 = 1$.
For the case with $p_8 =1$,
just three sets of SM fermions survive as zero modes such that
\begin{align}
&\psi^{1(2)}_L \supset (u_R^i)^c, (e_R^i)^c, q_L,~~
\psi^{2(1)}_R \supset d_R^i,(l_L)^c,
\nonumber \\
&\psi^{1(2)}_R \supset d_R, (l_L^i)^c,~~
\psi^{2(1)}_L \supset (u_R)^c, (e_R)^c, q_L^i, (\nu_R)^c,
\label{BC3}
\end{align}
where $i = 1, 2$.
For the case with $p_7 =1$,
just three sets of SM fermions survive as zero modes such that
\begin{align}
&\psi^{1(2)}_L \supset (u_R^i)^c, (e_R^i)^c, q_L,~~
\psi^{2(1)}_R \supset d_R, (l_L^i)^c,
\nonumber \\
&\psi^{1(2)}_R \supset d_R^i, (l_L)^c,~~
\psi^{2(1)}_L \supset (u_R)^c, (e_R)^c, q_L^i, (\nu_R)^c,
\label{BC4}
\end{align}
where $i = 1, 2$.

We summarize fermions with zero modes
and those gauge quantum numbers in Table \ref{Table:Property3&4}.
Here, $G_{322} = SU(3)_C \times SU(2)_L \times SU(2)_F$.

\begin{table}[htbp]
\begin{center}
  \caption{Gauge quantum numbers of fermions 
with even $Z_2$ parities for $SU(9) \rightarrow G_{322}
\times U(1)_1 \times U(1)_2 \times U(1)_3 \times U(1)_4$.}
  \label{Table:Property3&4}
\begin{tabular}{c|c|c|c|c|c|c|c}
\hline
species&$G_{322}$&$U(1)_1$&$U(1)_2$&$U(1)_3$&$U(1)_4$ 
& $p_8 = 1$ & $p_7 =1$ \\
\hline \hline
$(u_R^1)^c,(u_R^2)^c$&$(\overline{\bf 3},{\bf 1},{\bf 2})$&$-4$&$3$&$1$&$1$
& $\psi_L^{1(2)}$ & $\psi_L^{1(2)}$ \\
\hline
$(u_R)^c$&$(\overline{\bf 3},{\bf 1},{\bf 1})$&$-4$&$3$&$1$&$-2$
& $\psi_L^{2(1)}$ & $\psi_L^{2(1)}$ \\
\hline
$q_L^1,q_L^2$&$({\bf 3},{\bf 2},{\bf 2})$&$1$&$3$&$1$&$1$
& $\psi_L^{2(1)}$ & $\psi_L^{2(1)}$ \\
\hline
$q_L$&$({\bf 3},{\bf 2},{\bf 1})$&$1$&$3$&$1$&$-2$
& $\psi_L^{1(2)}$ & $\psi_L^{1(2)}$ \\
\hline
$(e_R^1)^c,(e_R^2)^c$&$({\bf 1},{\bf 1},{\bf 2})$&$6$&$3$&$1$&$1$
& $\psi_L^{1(2)}$ & $\psi_L^{1(2)}$ \\
\hline
$(e_R)^c$&$({\bf 1},{\bf 1},{\bf 1})$&$6$&$3$&$1$&$-2$
& $\psi_L^{2(1)}$ & $\psi_L^{2(1)}$ \\
\hline
$d_R^1,d_R^2$&$({\bf 3},{\bf 1},{\bf 2})$&$-2$&$-6$&$-2$&$1$
& $\psi_R^{2(1)}$ & $\psi_R^{1(2)}$ \\
\hline
$d_R$&$({\bf 3},{\bf 1},{\bf 1})$&$-2$&$-6$&$-2$&$-2$
& $\psi_R^{1(2)}$ & $\psi_R^{2(1)}$ \\
\hline
$(l_L^1)^c,(l_L^2)^c$&$({\bf 1},{\bf 2},{\bf 2})$&$3$&$-6$&$-2$&$1$
& $\psi_R^{1(2)}$ & $\psi_R^{2(1)}$ \\
\hline
$(l_L)^c$&$({\bf 1},{\bf 2},{\bf 1})$&$3$&$-6$&$-2$&$1$
& $\psi_R^{2(1)}$ & $\psi_R^{1(2)}$ \\
\hline
$(\nu_L)^c$&$({\bf 1},{\bf 1},{\bf 1})$&$0$&$-15$&$3$&$0$
& $\psi_L^{2(1)}$ & $\psi_L^{2(1)}$ \\
\hline
\end{tabular}
\end{center}
\end{table}

\section{Predictions}
\label{Sec:Predictions}

\subsection{Yukawa interactions}
\label{FM}

We examine whether four types of $SU(9)$ orbifold
family unification models,
where the embedding of the SM fermions are
realized as (\ref{BC1}), (\ref{BC2}), (\ref{BC3}) and (\ref{BC4}),
are realistic or not,
by adopting the appearance of Yukawa interactions 
from interactions in the 6D bulk as a selection rule.
This rule is not almighty to select models,
because Yukawa interactions can be constructed
on the fixed points of $T^2/Z_2$.
Here, we carry out the analysis 
under the assumption that such brane interactions
are small compared with the bulk ones in the absence of SUSY.

We assume that the Yukawa interactions in the SM
come from interaction terms
containing fermions in the bilinear form
and products of scalar fields in the 6D bulk.\footnote{
We assume that fermion condensations 
and Lorentz tensor fields are not involved
with the generation of Yukawa interactions.}
From the Lorentz, gauge and $Z_2$ invariance,
the Lagrangian density containing interactions
among a pair of Weyl fermions $(\Psi_+, \Psi_-)$ 
and scalar fields $\Phi^I$ on 6D space-time
is, in general, written as
\begin{align}
\mathcal{L}_{\rm int} &= \sum_{a,\cdots,f} 
\overline{\Psi}_{+abc} \Psi_-^{def} {F^{abc}}_{def}(\Phi^I)
+ \sum_{a,\cdots,f} \Psi_+^{Tabc} E \Psi_-^{def} G_{abcdef}(\Phi^I)
 + {\rm h.c.} \notag\\
&= \sum \left( \overline{\psi}^1_L\psi^1_R 
+ \overline{\psi}^2_R\psi^2_L \right) F(\Phi^I) 
+ \sum \left( (\psi^1_L)^{c\dagger}\psi^2_L 
+ (\psi^1_R)^{c\dagger}\psi^2_R \right) G(\Phi^I) + {\rm h.c.}, 
\label{Lint}
\end{align}
where $\overline{\Psi}_{+} \equiv \Psi_{+}^{\dagger}\Gamma^0$,
$\overline{\psi}_{L(R)}^{1(2)}=\psi_{L(R)}^{1(2)\dagger}\gamma^0$, 
and $(\psi_{L(R)}^{1(2)})^c=i\gamma^0\gamma^2\psi_{L(R)}^{1(2)\ast}$.
In the final expression of (\ref{Lint}),
we omit indices of $SU(9)$ such as $a$, $b$, $\cdots$, $f$
designating the components to avoid complications.
The $F(\Phi^I)$ and $G(\Phi^I)$ are some polynomials of $\Phi^I$,
e.g., $F(\Phi^I)$ is expressed by
\begin{align}
F(\Phi^I) = \sum_{I_1} f_{I_1} \Phi^{I_1} 
+ \sum_{I_1, I_2} f_{I_1 I_2} \Phi^{I_1} \Phi^{I_2} + \cdots
= \sum_n \sum_{I_1, \cdots, I_n} f_{I_1 \cdots I_n} \Phi^{I_1} \cdots \Phi^{I_n},
\label{F}
\end{align}
where $f_{I_1 \cdots I_n}$ are coupling constants.
Note that mass terms of $\Psi_{\pm}$ 
such as $m_{\rm D} \overline{\Psi}_{+} \Psi_{-}$
and $m_{\rm M} \Psi^{T}_{+} E \Psi_{-}$
are forbidden at the tree level,
in case that $\Psi_+$ and $\Psi_-$ have different intrinsic $Z_2$ parities.
Using the representation given by (\ref{Gammas}) and (\ref{gammas}),
$E$ is written as
\begin{align}
E \equiv \Gamma^1 \Gamma^3 \Gamma^6
= \left(\begin{array}{cccc} 
0 & 0 & i \sigma^2 & 0 \\
0 & 0 & 0 & i \sigma^2 \\
-i \sigma^2 & 0 & 0 & 0 \\
0 & -i \sigma^2 & 0 & 0 
\end{array} \right),
\end{align}
where $\sigma^2$ is the second element of Pauli matrices.
It is shown that $\mathcal{L}_{\rm int}$ is invariant 
under the 6D Lorentz transformation,
$\Psi_{\pm} \rightarrow 
\exp\left[-\frac{i}{4}\omega_{MN}\Sigma^{MN}\right]\Psi_{\pm}$,
where $\Sigma^{MN}=\frac{i}{2}[\Gamma^M,\Gamma^N]$ 
and $\omega_{MN}$ are parameters relating
6D Lorentz boosts and rotations. 

After the dimensional reduction occurs and
some components acquire the vacuum expectation values (VEVs)
generating the breakdown of extra gauge symmetries,
the linear terms of the Higgs doublet $\phi_h$ 
and its charge conjugated one $\tilde{\phi}_h$
can appear in $F(\Phi^I)$ and $G(\Phi^I)$
and then the Yukawa interactions are derived.
For instance, the linear term $\tilde{f} \phi_h$ appears
from $F(\Phi^I)= f \Phi_{1} \Phi_{3} \Phi_{5}$
where $\Phi_{m}$ are scalar fields whose representations
are ${9 \choose m}$, after some SM singlets in 
$\Phi_{3}$ and $\Phi_{5}$ acquire the VEVs.

From the above observations, we impose the selection rule that
{\it Yukawa interactions $f^u_{ij} \overline{q}_{L}^i u_{R}^j \tilde{\phi}_h$,
$f^d_{ij} \overline{q}_{L}^i d_{R}^j \phi_h$ 
and $f^e_{ij} \overline{l}_{L}^i e_{R}^j \phi_h$
in the SM
can be derived from $\mathcal{L}_{\rm int}$}
on orbifold family unification models.

For (BC1), the following Legrangian density is derived
at the compactification scale $M_{\rm C}$, 
\begin{align}
\mathcal{L}_{\rm (BC1)} 
= \sum_{i, j = 1}^3 \overline{d}_R^i q_L^j \tilde{F}_{1ij}^{(1)}(\phi)
+ \sum_{i, j = 1}^3 \overline{l}_L^i e_R^j \tilde{F}_{2ij}^{(1)}(\phi) 
+ \sum_{i, j = 1}^3 \overline{u}_R^i q_L^j \tilde{G}_{ij}^{(1)}(\phi) + {\rm h.c.},
\label{L-BC1}
\end{align}
using (\ref{BC1}),
and Yukawa interactions in the SM can be obtained,
after some SM singlet scalar fields in the
polynomials $\tilde{F}_{1}^{(1)}(\phi)$,  $\tilde{F}_{2}^{(1)}(\phi)$ 
and  $\tilde{G}^{(1)}(\phi)$ acquire the VEVs.
Because all gauge quantum numbers of the operator
$\overline{q}_L^i d_R^j$ are same as those of $\overline{l}_L^i e_R^j$,
there is a possibility that  $\tilde{F}_{1}^{(1)}(\phi)$ is 
identical with  $\tilde{F}_{2}^{(1)}(\phi)$ as a simple case.
In this case, we have the relations $f^d_{ij} = f^e_{ji}$ 
at the extra gauge symmetry breaking scale.

For (BC2), the following Legrangian density is derived, 
\begin{align}
\mathcal{L}_{\rm (BC2)} 
= \sum_{i, j = 1}^3 \overline{u}_R^i q_L^j \tilde{G}_{ij}^{(2)}(\phi) + {\rm h.c.},
\label{L-BC2}
\end{align}
using (\ref{BC2}).
In this case, down-type quark and charged leptons masses
cannot be obtained from $\mathcal{L}_{\rm int}$ 
at the tree level at $M_{\rm C}$.

For (BC3), the following Legrangian density is derived, 
\begin{align}
  \mathcal{L}_{\rm (BC3)} 
&= \sum_{i, j = 1}^2 \overline{d}_R^i q_L^j \tilde{F}_{1ij}^{(3)}(\phi)
+ \overline{q}_L d_R \tilde{F}_2^{(3)}(\phi)
+ \sum_{i, j = 1}^2 \overline{l}_L^i e_R^j \tilde{F}_{3ij}^{(3)}(\phi)
+ \overline{e}_R l_L \tilde{F}_4^{(3)}(\phi) + {\rm h.c.} \notag\\
&~~~~~~~~~~~
+ \sum_{i, j = 1}^2 \overline{u}_R^i q_L^j \tilde{G}_{1ij}^{(3)}(\phi) 
+ \overline{q}_L u_R \tilde{G}_2^{(3)}(\phi) + {\rm h.c.},
\end{align}
using (\ref{BC3}).
For (BC4), the following Legrangian density is derived, 
\begin{align}
  \mathcal{L}_{\rm (BC4)} 
&= \sum_{i =1}^2 \left(\overline{d}_R q_L^i \tilde{F}_{1i}^{(4)}({\phi})
+ \overline{q}_L d_R^i \tilde{F}_{2i}^{(4)}(\phi)
+ \overline{l}_L e_R^i \tilde{F}_{3i}^{(4)}(\phi)
+ \overline{e}_R l_L^i \tilde{F}_{4i}^{(4)}(\phi)\right) + {\rm h.c.} \notag\\
&~~~~~~~~~~~
+ \sum_{i, j = 1}^2 \overline{u}_R^i q_L^j \tilde{G}_{1ij}^{(4)}(\phi)
+ \overline{q}_L u_R \tilde{G}_2^{(4)}(\phi) + {\rm h.c.},
\end{align}
using (\ref{BC4}).
In both cases, the full flavor mixing cannot be realized 
at the tree level at $M_{\rm C}$.

In this way, we find that the model based on 
the embedding (\ref{BC1}) is a possible candidate
to realize the fermion mass hierarchy and flavor mixing,
in case that radiative corrections are too small to generate
mixing terms with suitable size for (BC2), (BC3) and (BC4).
In any case, we have no powerful principle to determine
the polynomials of scalar fields, and hence
we obtain no useful predictions from the fermion sector.

\subsection{Sfermion masses}
\label{SFM}

The SUSY grand unified theories on an orbifold
have a desirable feature that the triplet-doublet splitting
of Higgs multiplets is elegantly realized~\cite{YK,H&N}.
Hence, it would be interesting to construct
a SUSY extension of orbifold family unification models.

In the presence of SUSY, the model with (BC1) does not obtain
advantages of fermion sector over that with (BC2), (BC3) or (BC4),
because any interactions other than gauge interactions 
are not allowed in the bulk 
and Yukawa interactions must appear from brane interactions.
In SUSY models,  complex scalar fields 
$(\Phi_+, \Phi_-)$ are introduced 
as superpartners of $(\Psi_+, \Psi_-)$, and
they consist of two sets of complex scalar fields
$\Phi_+=(\phi_+^1, \phi_+^2)$ 
and $\Phi_-=(\phi_-^1, \phi_-^2)$,
where $\phi_+^1$, $\phi_+^2$, $\phi_-^1$ and $\phi_-^2$
are superpartners of $\psi_L^1$, $\psi_R^2$,
$\psi_R^1$ and $\psi_L^2$, respectively.
Here, we pay attention to 
superpartners of the SM fermions called sfermions
and study predictions of models.

Based on the assignment (\ref{BC1}) for (BC1),
sfermions are embedded into
scalar fields as follows,
\begin{align}
\phi_+^1 \supset \tilde{u}_R^{i\ast},~~\tilde{e}_R^{i\ast},~~\tilde{\nu}_R^{\ast},~~
\phi_+^2 \supset \tilde{d}_R^i,~~
\phi_-^1 \supset \tilde{l}_L^{i\ast},~~
\phi_-^2 \supset \tilde{q}_L^i.
\label{BC1-scalar}
\end{align}

Gauge quantum numbers for sfermions are given 
in Table $\ref{Table:SFPro}$. 
Here, the charge conjugation is performed
for scalar fields $\tilde{d}_R^i$ and $\tilde{l}_L^{i*}$
corresponding to the right-handed fermions,
and $G_{323} = SU(3)_C \times SU(2)_L \times SU(3)_F$.
Note that $(l_1,l_2,l_a)$ is untouched by change
as a mark of the place of origin in {\bf 84}.

\begin{table}[htbp]
\begin{center}
  \caption{Gauge quantum numbers of sfermions 
with even $Z_2$ parities for $SU(9) \rightarrow G_{323}
\times U(1)_1 \times U(1)_2 \times U(1)_3$.}
  \label{Table:SFPro}
\begin{tabular}{c|c|c|c|c|c}
\hline
species&$G_{323}$&$(l_1,l_2,l_a)$&$U(1)_1$&$U(1)_2$&$U(1)_3$ \\
\hline \hline
$\tilde{q}_L^i$&$({\bf{3}},{\bf{2}},{\bf{3}})$&$(1,1,1)$&$1$&$3$&$1$ \\
\hline
$\tilde{u}_R^{i\ast}$&$(\overline{\bf{3}},{\bf{1}},{\bf{3}})$&$(2,0,1)$&$-4$&$3$&$1$  \\
\hline
$\tilde{d}_R^{i\ast}$&$(\overline{\bf{3}},{\bf{1}},\overline{\bf{3}})$&$(1,0,1)$&$2$&$6$&$2$ \\
\hline
$\tilde{l}_L^i$&$({\bf{1}},{\bf{2}},\overline{\bf{3}})$&$(0,1,1)$&$-3$&$6$&$2$ \\
\hline
$\tilde{e}_R^{i\ast}$&$({\bf{1}},{\bf{1}},{\bf{3}})$&$(0,2,1)$&$6$&$3$&$1$ \\
\hline
$\tilde{\nu}_R^{\ast}$&$({\bf{1}},{\bf{1}},{\bf{1}})$&$(0,0,3)$&$0$&$-15$&$3$ \\
\hline
\end{tabular}
\end{center}
\end{table}

We study the sfermion masses based on the following two assumptions.\\
1) The SUSY is broken down by some mechanism
and sfermions acquire the soft SUSY breaking masses
respecting $SU(9)$ gauge symmetry.
Then, $\tilde{u}_R^{i\ast}$, $\tilde{e}_R^{i\ast}$, $\tilde{\nu}_R^{\ast}$
and $\tilde{d}_R^{i\ast}$ get a common mass $m_+$,
and $\tilde{q}_L^i$ and $\tilde{l}_L^i$ get a common mass $m_-$
at some scale $M_{\rm S}$.\\
2) Extra gauge symmetries 
$SU(3)_F \times U(1)_2 \times U(1)_3$ are broken down 
by the VEVs of some scalar fields at $M_{\rm S}$.
Then, the $D$-term contributions to the scalar masses
can appear as a dominant source of mass splitting.

The D-term contributions, in general, originate from $D$-terms 
related to broken gauge symmetries 
when the soft SUSY breaking parameters possess 
non-universal structure and the rank of gauge group decreases 
after the breakdown of gauge symmetry~\cite{KM&Y1,KM&Y2,D,H&K}. 
The contributions for scalar fields specifying by $(l_1,l_2,l_a)$
are given by
\begin{align}
  m^2_{D(l_1,l_2,l_a)} &= (-1)^{l_1 + l_2}
[Q_1 D_{F1} + Q_2 D_{F2} 
 + \{9(l_1+l_2)-15\} D_2  \notag\\
& ~~~~~~~~~~~~~~~~ + \{4 l_a-3(3-l_1-l_2)\} D_3],
\label{eq:sfmass}
\end{align}
where $Q_1$ and $Q_2$ are the diagonal charges 
(up to normalization) of $SU(3)_F$
for the triplet,
i.e., $(Q_1, Q_2) = (1, 1)$, $(-1, 1)$ and $(0, -2)$.
$D_{F1}$, $D_{F2}$, $D_2$ and $D_3$ are parameters 
including $D$-term condensations for broken symmetries. 

Using $m_+$, $m_-$ and $m^2_{D(l_1,l_2,l_a)}$, 
we derive the following formulae of mass square for each species 
at $M_{\rm S}$:\footnote{
In case that the extra gauge symmetry breaking scale ($M_{\rm F}$) is 
lower than $M_{\rm S}$,
$m_{\pm}^2$ receive radiative corrections between $M_{\rm S}$
and $M_{\rm F}$,
and the mass formulae should be modified.
Here, we consider the simplest case to avoid complications.
}
\begin{align}
  m^2_{\tilde{u}^{1\ast}_R} &= m_+^2+D_{F1}+D_{F2}+3D_2+D_3, \\
  m^2_{\tilde{u}^{2\ast}_R} &= m_+^2-D_{F1}+D_{F2}+3D_2+D_3, \\
  m^2_{\tilde{u}^{3\ast}_R} &= m_+^2-2D_{F2}+3D_2+D_3, \\
  m^2_{\tilde{e}^{1\ast}_R} &= m_+^2+D_{F1}+D_{F2}+3D_2+D_3, \\
  m^2_{\tilde{e}^{2\ast}_R} &= m_+^2-D_{F1}+D_{F2}+3D_2+D_3, \\
  m^2_{\tilde{e}^{3\ast}_R} &= m_+^2-2D_{F2}+3D_2+D_3, \\
  m^2_{\tilde{d}^{1\ast}_R} &= m_+^2-D_{F1}-D_{F2}+6D_2+2D_3, \\
  m^2_{\tilde{d}^{2\ast}_R} &= m_+^2+D_{F1}-D_{F2}+6D_2+2D_3, \\
  m^2_{\tilde{d}^{3\ast}_R} &= m_+^2+2D_{F2}+6D_2+2D_3, \\
  m^2_{\tilde{q}^{1}_{L}} &= m_-^2+D_{F1}+D_{F2}+3D_2+D_3, \\
  m^2_{\tilde{q}^{2}_{L}} &= m_-^2+D_{F1}-D_{F2}+3D_2+D_3, \\
  m^2_{\tilde{q}^{3}_{L}} &= m_-^2-2D_{F2}+3D_2+D_3, \\
  m^2_{\tilde{l}^{1}_{L}} &= m_-^2-D_{F1}-D_{F2}+6D_2+2D_3, \\
  m^2_{\tilde{l}^{2}_{L}} &= m_-^2-D_{F1}+D_{F2}+6D_2+2D_3, \\
  m^2_{\tilde{l}^{3}_{L}} &= m_-^2+2D_{F2}+6D_2+2D_3.
\end{align}

By eliminating unknown parameters such as $m_+^2$, $m_-^2$,
$D_{F1}$, $D_{F2}$, $D_2$ and $D_3$,
we obtain 15 kinds of relations\footnote{
Sum rules among sfermion masses have also been derived
using the orbifold family unification models 
on five-dimensional (5D) space-time~\cite{K&K1,K&K2,KK&M2}.
}
\begin{align}
  &m^2_{\tilde{u}^{1\ast}_R} = m^2_{\tilde{e}^{1\ast}_R},~~
m^2_{\tilde{u}^{2\ast}_R} = m^2_{\tilde{e}^{2\ast}_R},~~
m^2_{\tilde{u}^{3\ast}_R} = m^2_{\tilde{e}^{3\ast}_R}, \\
  &m^2_{\tilde{d}^{1\ast}_R}-m^2_{\tilde{l}^{1}_{L}}
=m^2_{\tilde{d}^{2\ast}_R}-m^2_{\tilde{l}^{2}_{L}}
=m^2_{\tilde{d}^{3\ast}_R}-m^2_{\tilde{l}^{3}_{L}} \notag\\
  &~~~~=m^2_{\tilde{u}^{1\ast}_R}-m^2_{\tilde{q}^{1}_{L}}
=m^2_{\tilde{u}^{2\ast}_R}-m^2_{\tilde{q}^{2}_{L}}
=m^2_{\tilde{u}^{3\ast}_R}-m^2_{\tilde{q}^{3}_{L}}, \\
  &m^2_{\tilde{q}^{1}_{L}}+m^2_{\tilde{l}^{1}_{L}}
=m^2_{\tilde{q}^{2}_{L}}+m^2_{\tilde{l}^{2}_{L}}
=m^2_{\tilde{q}^{3}_{L}}+m^2_{\tilde{l}^{3}_{L}}, \\
  &m^2_{\tilde{q}^{1}_{L}}+m^2_{\tilde{d}^{1\ast}_R}
=m^2_{\tilde{q}^{2}_{L}}+m^2_{\tilde{d}^{2\ast}_R}
=m^2_{\tilde{q}^{3}_{L}}+m^2_{\tilde{d}^{3\ast}_R} \notag\\
  &~~~~=m^2_{\tilde{l}^{1}_{L}}+m^2_{\tilde{u}^{1\ast}_R}
=m^2_{\tilde{l}^{2}_{L}}+m^2_{\tilde{u}^{2\ast}_R}
=m^2_{\tilde{l}^{3}_{L}}+m^2_{\tilde{u}^{3\ast}_R}.
\end{align}
They are compactly rewritten as
\begin{align}
  &m^2_{\tilde{u}^{i\ast}_R} = m^2_{\tilde{e}^{i\ast}_R},~~
m^2_{\tilde{d}^{i\ast}_R}-m^2_{\tilde{u}^{i\ast}_{R}}
=m^2_{\tilde{l}^{i}_L}-m^2_{\tilde{q}^{i}_{L}},\\
  &m^2_{\tilde{u}^{i\ast}_{R}}-m^2_{\tilde{u}^{j\ast}_{R}}
= -m^2_{\tilde{d}^{i\ast}_{R}}+m^2_{\tilde{d}^{j\ast}_{R}}
=m^2_{\tilde{q}^{i}_{L}}-m^2_{\tilde{q}^{j}_L}
=-m^2_{\tilde{l}^{i}_{L}}+m^2_{\tilde{l}^{j}_L},
\end{align}
where $i, j = 1, 2, 3$.

In the same way, 
based on (\ref{BC2}) for (BC2),
we obtain the relations,
\begin{align}
  &m^2_{\tilde{u}^{i\ast}_R} = m^2_{\tilde{e}^{i\ast}_R},~~
m^2_{\tilde{l}^{i}_L}-m^2_{\tilde{u}^{i\ast}_{R}}
=m^2_{\tilde{d}^{i\ast}_R}-m^2_{\tilde{q}^{i}_{L}},\\
  &m^2_{\tilde{u}^{i\ast}_{R}}-m^2_{\tilde{u}^{j\ast}_{R}}
= -m^2_{\tilde{d}^{i\ast}_{R}}+m^2_{\tilde{d}^{j\ast}_{R}}
=m^2_{\tilde{q}^{i}_{L}}-m^2_{\tilde{q}^{j}_L}
=-m^2_{\tilde{l}^{i}_{L}}+m^2_{\tilde{l}^{j}_L},
\end{align}
where $i, j = 1, 2, 3$.
Note that these relations are obtained by
exchanging $m^2_{\tilde{d}^{i\ast}_R}$ for 
$m^2_{\tilde{l}^{i}_{L}}$ in those for (BC1).

Furthermore, we obtain the specific relations,
\begin{align}
 &m^2_{\tilde{u}^{i\ast}_R} = m^2_{\tilde{e}^{i\ast}_R},~~
m^2_{\tilde{d}^{i\ast}_R}-m^2_{\tilde{u}^{i\ast}_{R}}
=m^2_{\tilde{l}^{i}_L}-m^2_{\tilde{q}^{i}_{L}},\\
&m^2_{\tilde{u}^{i\ast}_{R}}-m^2_{\tilde{u}^{j\ast}_{R}}
=-m^2_{\tilde{l}^{i}_{L}}+m^2_{\tilde{l}^{j}_L},~~
m^2_{\tilde{q}^{i}_{L}}-m^2_{\tilde{q}^{j}_L}
=-m^2_{\tilde{d}^{i\ast}_{R}}+m^2_{\tilde{d}^{j\ast}_{R}},\\
&m^2_{\tilde{u}^{1\ast}_{R}}-m^2_{\tilde{u}^{2\ast}_{R}}
=m^2_{\tilde{q}^{1}_{L}}-m^2_{\tilde{q}^{2}_L},\\
&m^2_{\tilde{u}^{1\ast}_{R}}+m^2_{\tilde{u}^{3\ast}_{R}}
=m^2_{\tilde{q}^{1}_{L}}+m^2_{\tilde{q}^{3}_L},~~
m^2_{\tilde{d}^{1\ast}_{R}}+m^2_{\tilde{d}^{3\ast}_{R}}
=m^2_{\tilde{l}^{1}_{L}}+m^2_{\tilde{l}^{3}_L}
\end{align}
for (BC3) and
\begin{align}
 &m^2_{\tilde{u}^{i\ast}_R} = m^2_{\tilde{e}^{i\ast}_R},~~
m^2_{\tilde{l}^{i}_L}-m^2_{\tilde{u}^{i\ast}_{R}}
=m^2_{\tilde{d}^{i\ast}_R}-m^2_{\tilde{q}^{i}_{L}},\\
&m^2_{\tilde{u}^{i\ast}_{R}}-m^2_{\tilde{u}^{j\ast}_{R}}
=-m^2_{\tilde{d}^{i\ast}_{R}}+m^2_{\tilde{d}^{j\ast}_{R}},~~
m^2_{\tilde{q}^{i}_{L}}-m^2_{\tilde{q}^{j}_L}
=-m^2_{\tilde{l}^{i}_{L}}+m^2_{\tilde{l}^{j}_L},\\
&m^2_{\tilde{u}^{1\ast}_{R}}-m^2_{\tilde{u}^{2\ast}_{R}}
=m^2_{\tilde{q}^{1}_{L}}-m^2_{\tilde{q}^{2}_L},\\
&m^2_{\tilde{u}^{1\ast}_{R}}+m^2_{\tilde{u}^{3\ast}_{R}}
=m^2_{\tilde{q}^{1}_{L}}+m^2_{\tilde{q}^{3}_L},~~
m^2_{\tilde{d}^{1\ast}_{R}}+m^2_{\tilde{d}^{3\ast}_{R}}
=m^2_{\tilde{l}^{1}_{L}}+m^2_{\tilde{l}^{3}_L}
\end{align}
for (BC4).
Here, $i, j = 1, 2, 3$
and we denote $\tilde{u}^{\ast}_R$, $\tilde{e}^{\ast}_R$,
$\tilde{d}^{\ast}_R$, $\tilde{l}_L$ and $\tilde{q}_L$
as $\tilde{u}^{3\ast}_R$, $\tilde{e}^{3\ast}_R$, $\tilde{d}^{3\ast}_R$,
$\tilde{l}^{3}_L$ and $\tilde{q}^{3}_L$.
The relations for (BC4) are obtained by
exchanging $m^2_{\tilde{d}^{i\ast}_R}$ for 
$m^2_{\tilde{l}^{i}_{L}}$ in those for (BC3).

The above relations become predictions to probe models
because they are specific to models,
in case that the extra gauge symmetry breaking scale is near $M_{\rm S}$.

\section{Conclusions and discussions}
\label{C&D}

We have taken orbifold family unification models
based on $SU(9)$ gauge symmetry 
on $M^4 \times T^2/Z_2$ as a starting point
and have examined the reality of models, 
by adopting the appearance of Yukawa interactions
from the interactions in the 6D bulk as a selection rule.
We have picked out a candidate of model 
compatible with the observed fermion masses and flavor mixing.
The model has an feature that just three families of
fermions in the SM exist as zero modes
and any mirror particles do not appear in the low-energy world
after the breakdown of gauge symmetry
$SU(9) \rightarrow SU(3)_C \times SU(2)_L 
\times SU(3)_F \times U(1)^3$ by orbifolding.
Depending on the assignment of intrinsic $Z_2$ parities,
$((u_R^i)^c, (e_R^i)^c, d_R^i)$ and $((l_L^i)^c, q_L^i)$ belong to 
$\Psi_{\pm}$ and $\Psi_{\mp}$ with {\bf 84} of $SU(9)$, respectively.
We have found out specific relations
among sfermion masses 
as model-dependent predictions
in the SUSY extension of models.

The mass degeneracy for each squark and slepton species 
in the first two families is favorable 
for suppressing flavor-changing neutral current (FCNC) processes. 
The $D$-term contributions relating $SU(3)_F$, however,
can spoil the mass degeneracy.
Such dangerous situations induce sizable FCNC processes
can be avoided 
if the sfermion masses in the first two families are rather large, 
the fermion and its superpartner mass matrices are aligned,
or the $D$-term contributions to lift the degeneracy are small enough.

As a future work, we need to answer the question
whether the fermion mass spectrum 
and flavor mixing are successfully achieved 
at the low energy scale, in our orbifold family unification model.

\section*{Acknowledgments}
This work is supported in part by funding from Nagano Society for The Promotion of Science (Y. Goto).

\end{document}